\documentclass[preprint,12pt]{elsarticle}

\usepackage{amssymb}

\usepackage{booktabs} 
\usepackage{tabularx} 
\newcommand{\tableheadline}[1]{\multicolumn{1}{l}{\textbf{#1}}}
\usepackage{amssymb}
\usepackage{xcolor}
\usepackage{amsmath}
\usepackage{float}
\usepackage{amsfonts}
\usepackage{amssymb}
\usepackage{amsthm}
\usepackage{makeidx}
\usepackage{amsmath}
\usepackage{amsfonts}
\usepackage{amssymb}
\usepackage{graphicx}
\usepackage{afterpage}
\usepackage{pdfpages}
\usepackage{comment}
\usepackage{subcaption}

\usepackage[colorlinks=true,linktocpage=true,pagebackref=true, citecolor=red,linkcolor=blue]{hyperref}

\newtheorem{thm}{Theorem}[section]
\newtheorem{prop}[thm]{Proposition}
\newtheorem{defn}[thm]{Definition}

\newcommand{\add}{\vspace{8pt}\noindent\addtocounter{thm}{1}}
\newcommand{\example}{\add\noindent{\bf Example \thethm \quad }}

\newcommand{\remark}{\add\noindent{\bf Remark \thethm \quad}}
\newcommand{\remarks}{\add\noindent{\bf Remarks \thethm \quad}}

\journal{arXiv}

\begin{document}
	
	\begin{frontmatter}
		
		
		
		\title{The logistic queue model: theoretical properties and performance evaluation}
		
		
		\author[mymainaddress,mysecondaryaddress]{Franco Coltraro\corref{mycorrespondingauthor}}
		\cortext[mycorrespondingauthor]{Corresponding author}
		\ead{franco.coltraro@upc.edu}
		
		\author[mysecondaryaddress]{Marc Ruiz}
		\author[mysecondaryaddress]{Luis Velasco}

		\address[mymainaddress]{Institut de Rob\`otica i Inform\`atica Industrial, CSIC-UPC, Barcelona, Spain.}
		\address[mysecondaryaddress]{Universitat Polit\`ecnica de Catalunya $\cdot$ BarcelonaTech, Barcelona, Spain.}

		\begin{abstract}
			The advent of digital twins (DT) for the control and management of communication networks requires accurate and fast methods to estimate key performance indicators (KPI) needed for autonomous decision-making. Among several alternatives, queuing theory can be applied to model a real network as a queue system that propagates entities representing network traffic. By using fluid flow queue simulation and numerical methods, a good trade-off between accuracy and execution time can be obtained. In this work, we present the formal derivation and mathematical
			properties of a continuous fluid flow queuing model called the logistic queue model.
			We give novel proofs showing that this queue model has all the theoretical properties one should expect such as positivity of the queue and first-in first-out (FIFO) property. Moreover, extensions are presented in order to model different characteristics of telecommunication networks, including finite buffer sizes and propagation of flows with different priorities. Numerical results are presented to validate the accuracy and improved performance of our approach in contrast to traditional discrete event simulation, using synthetic traffic generated with the characteristics of real captured network traffic. Finally, we evaluate a DT built using a queue system based on the logistic queue model and demonstrate its applicability to estimate KPIs of an emulated real network under different traffic conditions.
			
		\end{abstract}
		
		
		
		
		
		
	\end{frontmatter}
	
	
	\section{Introduction}
	
	Over the past decade, network communication systems have transformed into a fundamental infrastructure that supports digital demands from all industry sectors \cite{Uu21}. The advent of beyond 5G (B5G) and 6G paradigms is expected to bring heterogeneous networks providing smart end-to-end connectivity to a plethora of extremely variate devices, supporting entirely diverse classes of services with outstanding performance while making the communication infrastructure fully transparent to the applications \cite{Ba21}. Among different network technologies, the most relevant include: 6th generation of cellular radio access networks \cite{Po23}, ultra-high capacity optical networks \cite{Ru23.2}, adaptive packet networks consisting of programmable packet nodes \cite{Pa19}; and deterministic networks aiming at delivering time-sensitive networking services requiring predictable latency-based performance \cite{Sa23}.
	
	\medskip
	
	Managing these complex and heterogenous networks requires from high degree of automation, which entails the adoption of advanced artificial intelligence (AI)-based mechanisms \cite{Bt20}. Among different technologies and trends, digital twins (DT) are attracting large attention and concentrating remarkable research effort. In the context of telecommunication networks, a DT is a virtual representation of a network segment or domain that is used for predictive performance analysis and network diagnosis \cite{De24}. Thus, in order to efficiently manage the connectivity of end-to-end flows in support of B5G and 6G services, DTs are required for flow performance estimation in terms of key indicators such as throughput and delay \cite{Be20}.
	
	\medskip
	
	To achieve that goal, we recently developed a DT called CURSA-SQ that consists in a set of flow traffic generators, fluid flow queue models, and algorithms that enable a virtual representation of the network system for performance evaluation purposes. The initial models of CURSA-SQ were presented in \cite{Ru18}, where the logistic queue model for flow traffic analysis in fixed transport networks was introduced for the first time. Moreover, we presented initial extensions of that logistic queue model for modeling wireless network segments \cite{Be20} and time-sensitive network interfaces \cite{Ve20}. However, none of the aforementioned publications present details about how the equations were derived or the mathematical proofs supporting the validity of the queue model. In addition, they focus on particular technologies, which does not demonstrate its application to a broader range of B5G and 6G systems and services expected to come.
	
	\medskip
	
	In this paper, we cover the aforementioned lacks and issues in previous works, and focus on the formal definition of the logistic queue model, highlighting its benefits and differences with respect to existing fluid flow models. The logistic model is presented as a general purpose, technology agnostic model, with a set of proven properties such as queue positiveness and first-in first-out (FIFO) dynamic, and a list of extensions such as finite queue size, time-dependent server rate and priority-based flow management, that allows its application to the wide range of envisioned B5G and 6G systems.
	
	\medskip
	
	Complementing the formal definition and mathematical proofs, a set of testing results are presented in order to evaluate the accuracy and validity of the logistic queue model to be applied in a DT context. This includes a detailed evaluation of metrics such as queue buffering and outflow dynamics, which are essential to estimate key performance indicators such as delay and throughput. Finally, execution time is an important aspect since typically DTs operate in near-real time conditions ($<1$-$5$ seconds). Thus, execution times are provided, in comparison with other approaches such as discrete-event simulation.

	\section{Related work}
	
	As already mentioned, the optimization of telecommunication cloud infrastructures includes network planning and reconfiguration by means of various mathematical and computational techniques \cite{rutas}. Among them, in the context of DTs, dynamic network simulation is of paramount importance in order to evaluate the performance of new services and applications over the network. To that end, one of the key elements in the simulation is how to generate and model network traffic. Traffic generation is a useful technique that enables studying and evaluating the network performance through simulation, when, e.g. real traffic traces are not available. 
	
	\medskip
	
	Simulations based on flow-based models, e.g. \cite{alba}, present interesting features such as optimum efficiency and easiness of traffic parameter characterization. With such approaches, traffic generators are developed to inject realistic traffic flows in telecom cloud-simulated systems. Basically, different types of functions like piece-wise linear, polynomial or trigonometric sine summation are used to model traffic flow average profiles that can be periodical and evolutionary, e.g. incremental. Besides, traffic is characterized by a random function around that average. However, the mix of services that could be aggregated into a flow could be too heterogeneous and variable to allow a good characterization following such a simple approach. Moreover, interesting outputs that could be measured in simulation such as end-to-end latency or node switching delay cannot be accurately obtained due to the inherent nature of flow-based models that hides any individual service behavior.

	\medskip
	
	An alternative approach is to generate and simulate traffic at the packet level, i.e. characterizing the packet generation of an individual service and running a discrete event simulation that processes the transition of the packet from source to destination to every intermediate node \cite{omnet,discrete_queue}. It is worth noting that the amount of information that could be obtained from a discrete event simulation based on a packet-based traffic generator is unbeatable. However, when high-income bitrates e.g. in the order of dozens of Gb/s per flow, the computational cost of processing the resulting huge number of packets is prohibitive even for small networks. To this aim, in the last years several research works have focused on providing alternative simulation environments allowing a fine granular view of the system comparable with that of packet-based simulations with the efficiency and scalability of flow-based ones.
	
	\medskip
	
	One of the most successful approaches for doing so has been the use of \textit{fluid flow models} \cite{fluid_flow}. Basically, the idea is
	to consider only changes in rates of traffic flows. This can
	result in large performance advantages, though information
	about the individual packets is lost. For this reason, also hybrid models have been developed \cite{hybrid}. In this type of mixed models, a fluid simulator records
	the changes in the fluid rate in the source and the queue,
	while a packet simulator records the events of all the
	packets in the system. The abstraction takes place when packet flows with little \textit{time slots} separations are considered to be in the same fluid
	flow with a constant fluid rate. Little time variations
	among packets are not considered, and in this way, the
	number of events is reduced. Of course, a critical issue of this kind of models is how to choose the time slots when one is given an arbitrary flow. Moreover, recording the changes in the fluid rate can still be seen as being discrete in nature. 
	
	\medskip
	
	In the end, that type of models can be seen as discrete variations of the famous \textit{Vickrey's point-queue model} \cite{cola_puntual}. This is a purely \textit{continuous} model --we do not have to worry about time slots-- formulated through a differential equation. It has been thoroughly studied, being one of the most notable works the one done in \cite{pde1}.  In that article, the authors find an explicit solution to the previous differential equation and with that formula at hand, they are able to prove many desirable properties of the point-queue model, e.g. positivity of the queue size and the FIFO property. Nevertheless, there are some issues with the model. The most obvious one is that the right-hand side of the differential equation is not continuous and hence in general there do not exist classical solutions. This is computationally and numerically problematic since we can not apply usual ordinary differential equation (ODE) integrators. Nevertheless, it is worth noting that the authors give in \cite{pde2} an easily applicable numerical algorithm. The main problem that remains is generalization. Since the numerical algorithm they derive relies on the exact solution they found, we can not expect to have numerical schemes for more general cases as finite queues or priority queues.
	
	\subsection{Contributions}
	
	In this work we present for the first time a formal derivation of the \textit{logistic queue model},  addressing on the way the aforementioned problems. We explain how this model can be seen as a smooth formulation of Vickrey's point-queue model. This is important since in such a way we are able to use usual numerical integrators. We prove mathematically that the model has all the theoretical properties one should expect: i) \textit{positivity} of the queue size, ii) \textit{asymptotic behavior}: the queue gets empty if the inflow does not overflow and iii) \textit{FIFO property}: the system satisfies a first in, first out discipline.
	
	\medskip
	
	Moreover, we validate the logistic queue model comparing it with a discrete event simulator. This way we show that for many purposes this model is as precise as a discrete one with the advantage of speed in simulations. We compare simulation times and conclude that the logistic queue model is several orders of magnitude faster than a discrete one. Finally, in contrast with the point-queue model, the logistic one allows us to easily explore multiple extensions to more general scenarios such as finite queues (see Theorem \ref{main_thm}), multiple servers, priority queues, etc.
	
	\subsection{Organization}
	
	The rest of this work is organized as follows: in Section \ref{sec:derivation} we present the logistic queue model. There, we give a full derivation of it, with all the motivations behind the equations. In Section \ref{sec:proofs} we present original proofs of theoretical properties of the model. In Section \ref{sec:extensions} we discuss important extensions of the model so that it can handle more general situations common in telecommunication networks. In Section \ref{sec:method} we explain how to set up a discrete packet simulator in order to have a benchmark to compare the logistic queue model with. In Section \ref{sec:validation} we validate the model by comparing its performance with the discrete event simulator. Finally, in Section \ref{sec:scenario} we give an application of the model to show its potential. The article ends with some concluding remarks.
	
	\subsection{Notations}
	In this section, we make a summary of all our notation and its meaning. They will be used thoroughly in the following sections. 
	\begin{enumerate}
		\item[-] \textit{Maximum outflow rate}: $\mu$. Its units are  entities/time. It will be assumed to be constant, for example, $\mu = 1$ Gb/s. It is given, so that it is an input.
		\item[-] \textit{Queue size at $t$}: $q(t)$. Its units are entities. It varies over time. Our goal is to predict this quantity.
		\item[-] \textit{Maximum queue capacity}: $k$. Its units are entities. It is fixed in time. This is an input of the system.
		\item[-] \textit{Inflow to the system at $t$}: $X(t)$. Its units are entities/time. It varies over time. It is given, so that it is an input.
		\item[-] \textit{Outflow of the system at $t$}: $Y(t)$. Its units are entities/time. It varies over time. Our goal is to predict this quantity.
		\item[-] \textit{Mean inflow}: $\lambda$. Its units are entities/time. It is the mean of the inflow $X$.
		\item[-] \textit{Mean intensity or occupancy}: $\rho$. It has no units. It is defined as $\tfrac{\lambda}{\mu}$. This quantity gives us an idea of the average use of the server.
		\item[-] \textit{Aggregation time}: $dt$. Its units are time. When a flow is given in discrete form, we will assume it is given each $dt$ seconds. Usually $dt = 60$.
		\item[-] \textit{Logistic model parameter}: $\alpha$. It is defined as $\tfrac{\rho}{\mu}$.
	\end{enumerate} 
	
	\section{Derivation of the logistic queue model}\label{sec:derivation}
	
	In this section, we derive the logistic queue model. We show how fluid flow models come up as a consequence of a conservation law. Then the logistic queue model appears as a functional relationship of the queue size, the inflow and the outflow.
	
	\smallskip
	
	Assume we have a system with an inflow of entities arriving to a server of constant speed $\mu$ with an infinitely long queue. In probabilistic terminology, the system is denoted as $G/D/1$. We will derive an equation that relates the number of entities in queue $q$ with the inflow to the system $X$ and the outflow $Y$.

	\begin{figure}[h!]
		\begin{center}
			\includegraphics[scale=0.35]{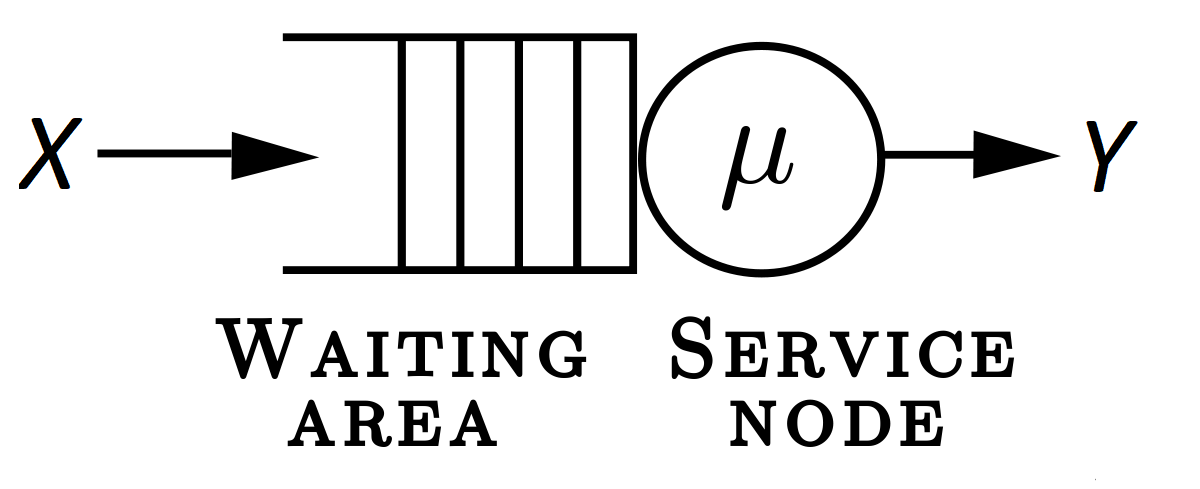}
			\caption{Example of a system with a queue and a server of speed $\mu$. }
		\end{center}
	\end{figure}
	\noindent Note that each quantity has the following units:
	\begin{enumerate}
		\item[1.] $[\mu] = $ entities/time.
		\item[2.] $[q] = $  entities.
		\item[3.] $[X] = [Y] = $ entities/time.
	\end{enumerate}
	We will assume that the amount of entities in queue $q(t)$ in the instant $t$ is a real number. This will be a reasonable approximation if the inflow is large enough: $X\gg1$, physically this means that \textit{many} entities arrive to the system. Also, this will make us not to distinguish between entities in queue and entities in the system. Therefore, for a short period of time $dt$ the amount of entities that have arrived to the system is approximately $X(t)\cdot dt$ and likewise the number of them that have abandoned it is $Y(t)\cdot dt$. This gives us the following conservation law:
	\begin{equation*}
		q(t+dt) = q(t) +  X(t)\cdot dt - Y(t)\cdot dt,
	\end{equation*}
	which in the limit when $dt\rightarrow0$ reads:
	\begin{equation}\label{conservation}
		q'(t) = X(t) - Y(t).
	\end{equation}
	\subsection{Outflow.} 
	We shall assume that there is a functional relationship between $X,Y$ and $q$, and express the \textbf{outflow} as a function of the inflow and the queue size:
	\begin{equation}\label{out}
		Y(t) = Y(X,q) = \mu + e^{-\alpha q(t)}\left[\min\{\mu,X(t)\}-\mu\right].
	\end{equation} 
	where:
	
	\begin{enumerate}
		\item[1.] $\mu>0$ is maximum outflow rate of the server.
		\item[2.] $\alpha$ is a positive parameter taken to be $\tfrac{\rho}{\mu}$ where $\rho = \tfrac{\lambda}{\mu}$ is the average intensity,
		\begin{equation*}
			\lambda = \tfrac{1}{t_1-t_0}\int_{t_0}^{t_1}X(t)dt
		\end{equation*}
		is the mean inflow and $[t_0,t_1]$ is the interval of definition of $X$.
	\end{enumerate}
	
	\remarks Note that by construction we have:
	\begin{enumerate}
		\item[1.] If at $t$ the queue is empty, i.e. $q(t)=0$, then the outflow is just $\min\{\mu,X(t)\}$.
		\item[2.] If the inflow is bigger than the maximum capacity, i.e. $X(t)>\mu$, then the outflow is just $\mu$.
		\item[3.] If $q(t)\rightarrow+\infty$ then the outflow tends to $\mu$.
	\end{enumerate}
	
	\noindent Finally we put together (\ref{conservation}) and (\ref{out}) in the following definition:
	\begin{defn}[\textbf{Logistic queue model}]
		\em Given an initial condition $q(t_0)=q_0$, a continuous and positive inflow function $X:[t_0, +\infty)\rightarrow [0, +\infty)$ and a maximum outflow rate $\mu>0$, the \textit{logistic queue model} is the following ordinary differential equation:

		\begin{equation}\label{modlog}
			\begin{cases}
				q'(t) &= X(t) - \left[\mu + e^{-\alpha q(t)}\left(\min\{\mu,X(t)\}-\mu\right)\right] \text{  for  } t>t_0,\\
				q(t_0) &= q_0.
			\end{cases}
		\end{equation}	
	\end{defn}
	\remark If we let $\alpha\rightarrow+\infty$ we get the famous \textit{Vickrey's point-queue model}:
	\begin{equation}\label{modpoint}
		q'(t) = X(t) -  \begin{cases}
			\min\{\mu,X(t)\}& \text{  if  } q(t)= 0,\\
			\mu& \text{  if  } q(t)\neq0.
		\end{cases}   
	\end{equation}
	Note that the right-hand side of the ODE is not continuous in $q$. This causes many computational and theoretical difficulties. In the next section, we will prove that the logistic queue model has all the good properties of the point-queue model discussed in the introduction with the additional advantage that it is easy to integrate numerically because the ODE (\ref{modlog}) is smooth.
	
	\section{Mathematical properties of the logistic queue model}\label{sec:proofs}
	In this section, we present the main theoretical properties of the logistic queue model. We give original mathematical proofs of reasonable properties all queueing models ought to have: positivity of the queue, emptying of the queue and FIFO property. Let's start with existence, uniqueness and positivity. 
	
	\begin{prop}[\textbf{Existence, Uniqueness and Positivity}]\label{positividad}
		Given an initial condition $q(t_0)=q_0\geq 0$, a continuous and positive inflow function $X\geq0$ and a maximum outflow rate $\mu>0$, we have that there exists a unique continuous differentiable solution $q(t)$ to the system (\ref{modlog}) defined in an interval $[t_0,T_{max})$ for $T_{max}\leq+\infty$. Moreover, such a solution is always greater or equal to zero. 
	\end{prop}
	
	\begin{proof}
		Existence, uniqueness and differentiability of $q(t)$ in an interval $[t_0,T_{max})$ for $T_{max}\leq+\infty$ is an easy consequence of the fact that the right-hand side of (\ref{modlog}) is a smooth function of the variable $q$ and it is continuous in $t$, hence we can apply \textit{Cauchy-Lipschitz theorem} \cite{ODEth}. Let us prove now positiveness. Assume that for some time $t_{-}>t_0$ we have that $q(t_{-})<0$. Then, since $q(t)$ is continuous and $q_0\geq0$, there must be a $t_{+}$ and a $t_{*}$ such that $t_{+}<t_{*}<t_{-}$ where $q(t_{+})\geq0$ and $q(t_{*})=0$. Also, after reducing the interval $I = [t_{+},t_{-}]$ if necessary, we may assume that $q(t)$ is strictly decreasing there. Therefore we have:
		
		\begin{equation*}
			q'(t)\leq 0 \text{ in } I = [t_{+},t_{-}].
		\end{equation*}

		Let's see how this implies that $X(t)\leq\mu$ in $I$. Indeed if $X(t)>\mu$ for some $t\in I$ we would have using (\ref{modlog}) that $q'(t) = f(t)- \mu > 0$, which is a contradiction. Hence we deduce that $\tilde{q}(t)\equiv0$ is a solution of
		\begin{equation}\label{modlogloc}
			\begin{cases}
				q'(t) &= X(t) - \left[\mu + e^{-\alpha q(t)}\left(\min\{\mu,X(t)\}-\mu\right)\right]  \text{ for } t\in I,\\
				q(t_*) &= 0.
			\end{cases}
		\end{equation}
		
		This is impossible because we would have two different solutions $q$ and $\tilde{q}\equiv0$ of (\ref{modlogloc}) in $I$.
	\end{proof}

	\remark As a consequence of the previous proposition we deduce that if the queue starts empty, i.e. $q(t_0)=0$, and the inflow is always smaller than the maximum outflow rate: 
	\begin{equation*}
		X(t)<\mu\text{ for all }t\geq t_0,
	\end{equation*}
	then the queue remains empty for all times and the outflow is equal to the inflow.
	
	\bigskip
	
	Now we are ready to prove some asymptotical properties of the queue:
	
	\begin{prop}[\textbf{Asymptotical behaviour}]
		Let $q(t_0)=q_0\geq 0$ be an initial condition, $X\geq 0$ a continuous and positive inflow function and $\mu>0$ a maximum outflow rate. Then the solution of (\ref{modlog}) is defined for all times, i.e. it is defined in the interval $[t_0,+\infty)$. Moreover if the inflow satisfies: 
		\begin{equation}\label{asymp}
			X(t)\leq X_{\infty}<\mu \text{ for all } t\geq t_X
		\end{equation}
		for some time $t_X>t_0$ then 
		\begin{equation*}
			\lim_{t\rightarrow+\infty} q(t) = 0
		\end{equation*}
		exponentially fast.
	\end{prop}
	
	\begin{proof}
		Note that we can rewrite (\ref{modlog}) as:
		\begin{equation}\label{modlogv2}
			q'(t) = X(t) - \mu + e^{-\alpha q(t)}\left(\mu - \min\{\mu,X(t)\}\right),
		\end{equation} 
		but since $q(t)\geq 0$ we have that $e^{-\alpha q(t)}\leq1$, and moreover $\mu - \min\{\mu,X(t)\}\leq \mu$. Hence we deduce that: 
		\begin{equation*}
			q'(t) \leq X(t) - \mu + \mu = X(t),
		\end{equation*} 
		so finally we get that:
		\begin{equation*}
			q(t) \leq q_0 + \int_{t_0}^{t}X(s)ds.
		\end{equation*}
		Therefore $q(t)$ is defined for all times since $X$ is continuous and hence integrable. Assume now that $X$ satisfies (\ref{asymp}). Then, using (\ref{modlogv2}), we get:
		\begin{align*}
			q'(t) &= (1 - e^{-\alpha q(t)})\cdot(X(t) - \mu) \\
			&\leq (1 - e^{-\alpha q(t)})\cdot(X_{\infty} - \mu), \quad \text{for all } t\geq t_X.
		\end{align*} 
		Now, it is well known that the exponential satisfies:
		\begin{equation*}
			e^{\alpha q} = \sum_{n = 0}^{\infty}\frac{(\alpha q)^n}{n!} \geq 1 + \alpha q,\quad \text{for all } q\geq 0.
		\end{equation*}
		Hence,
		\begin{equation*}
			e^{-\alpha q} \leq \frac{1}{1 + \alpha q} \Leftrightarrow  1 - e^{-\alpha q} \geq 1 - \frac{1}{1 + \alpha q} =  \frac{\alpha q}{1 + \alpha q}.
		\end{equation*}
		Now, since $(X_{\infty} - \mu)<0$ we get the following inequality:
		\begin{equation*}
			q' \leq \frac{\alpha q\cdot(X_{\infty} - \mu)}{1 + \alpha q}.
		\end{equation*} 
		Calling $\beta = \mu - X_{\infty} > 0$ and rearranging terms we obtain:
		\begin{equation*}
			(1 + \frac{1}{\alpha q})\cdot q' \leq -\beta \Leftrightarrow \frac{d}{dt}\left(q + \frac{\log q}{\alpha}\right) \leq \frac{d}{dt}\left(-\beta t\right).
		\end{equation*} 
		Therefore, integrating from $t_X$ to $t$,
		\begin{equation*}
			q(t) + \frac{\log q(t)}{\alpha} \leq -\beta t + k,\quad \text{for all } t\geq t_X
		\end{equation*} 
		where $k = q_X + \frac{\log q_X}{\alpha} +\beta t_X$ and $q_X = q(t_X)$. Applying the exponential in both sides of the inequality one gets:
		\begin{equation}\label{bound_queue}
			q^{1/\alpha}\cdot e^{q} \leq e^{-\beta t + k}.
		\end{equation}
		Finally, the last equation implies that
		\begin{equation*}
			q(t) \leq e^{\alpha(k -\beta t )},\quad \text{for all } t\geq t_X.
		\end{equation*}
		So the queue gets empty exponentially fast as stated. 
	\end{proof}
	\remark Equation (\ref{bound_queue}) gives us an estimate of how fast the queue goes to zero. Assuming (\ref{asymp}) and given a fixed tolerance $\epsilon>0$, if we impose that:
	\begin{equation*}
		q(t) \leq e^{\alpha(k -\beta t )}\cdot e^{-\alpha q(t)}\leq \epsilon,
	\end{equation*}
	then we find that the \textbf{emptying time} $T_\epsilon(q_X)\geq t_{X}$ needed to empty the queue from $q_X$ within a tolerance $\epsilon>0$ satisfies the following bound:
	\begin{equation}
		T_\epsilon(q_X) \leq t_X + \frac{q_X-\epsilon}{\mu - X_{\infty}} + \frac{1}{\alpha\cdot(\mu - X_{\infty})}\cdot \log\left(\frac{q_X}{\epsilon}\right).
	\end{equation}
	Note that the previous equation has the following expected \textit{physical} properties:
	\begin{enumerate}
		\item[1.] The fastest way to empty the queue is by stopping the inflow, i.e. taking $X_\infty = 0$. On the other hand, if $X_\infty \rightarrow\mu$ then the bound goes to infinity as expected.
		\item[2.] If $\alpha \rightarrow+\infty$ then we get that the bound is equal to the one of the \textit{point-queue model.} 
		\item[3.] The bound is optimal because if we take $\epsilon = q_X$ then we deduce that $T_\epsilon(q_X) = t_X$.
	\end{enumerate}
	To end this section we will show that our model satisfies the FIFO (first-in first-out) property, this is, that the entity that arrives first to the queue is also the one that gets out first. To accomplish this, we introduce the \textbf{exit time} of an entity that has arrived in the instant $t$ as:
	\begin{equation}
		\Lambda(t) = t + \frac{q(t)}{\mu}.
	\end{equation} 
	
	\remark The rationale behind this definition is the following: $q(t)/\mu$ has units of time and it represents approximately the amount of time needed for emptying the queue if no more entities arrived after $t$.
	
	\smallskip
	
	We have the following proposition:

	\begin{prop}[\textbf{FIFO property}]
		Let $q(t_0)=q_0\geq 0$ be an initial condition, $X\geq 0$ a continuous and positive inflow function and $\mu>0$ a maximum outflow rate. Then 
		\begin{equation*}
			\text{ If } t<s \text{ we have that } \Lambda(t)<\Lambda(s). 
		\end{equation*}
		In other words, the exit time of an entity that has arrived in the instant $t$ is smaller than others that arrive at $s>t$.
	\end{prop}
	\begin{proof}
		We have that $\Lambda(t)<\Lambda(s)$ is equivalent to:
		\begin{equation} \label{fifo}
			\frac{q(s)-q(t)}{s-t}> -\mu.
		\end{equation}
		Since $q$ is differentiable, by the \textit{Mean-Value theorem} we know that there exists a $\xi\in(t,s)$ such that
		\begin{equation*}
			q'(\xi) = \frac{q(s)-q(t)}{s-t}.
		\end{equation*}
		Now, we distinguish two cases:
		\begin{enumerate}
			\item[1.] $X(\xi)<\mu$. There are two sub-possibilities:
			\medskip
			\begin{enumerate}
				\item[1.1] $q(\xi) = 0$. In this case $q'(\xi) = X(\xi)-X(\xi) = 0 > -\mu$, so (\ref{fifo}) holds.
				\smallskip
				\item[1.2] $q(\xi) > 0$. In this case the outflow $g$ satisfies $0<g(\xi)<\mu$, hence
				\begin{equation*}
					q'(\xi) = X(\xi)-g(\xi)\geq-g(\xi)>-\mu.
				\end{equation*}
			\end{enumerate}
			\item[2.] $X(\xi)\geq\mu$. In this case $q'(\xi) = X(\xi) - \mu > -\mu$, so (\ref{fifo}) holds.
		\end{enumerate}
	\end{proof}

	\section{Logistic queue model extensions}\label{sec:extensions}
	In this section we present variations of the basic logistic queue model that allow us to address more general scenarios present in telecommunication networks. Namely:
	\begin{enumerate}
		\item[$\bullet$]  The case in which we have a finite queue.
		\item[$\bullet$]  The case in which we have a probabilistic distribution on service times, i.e. a G/G/1 system.
		\item[$\bullet$]  The case in which we have $m$ servers, i.e. a G/D/m system.
		\item[$\bullet$]  The case in which two flows join in a server and then separate following different paths.
		\item[$\bullet$]  The case in which we have priority queues.
	\end{enumerate} 
	Note that we may also have combinations of the previous cases.
	\subsection{{Finite queue}}
	Assume we are only able to store a finite number of entities $k>0$ in queue. We would like to add this restriction to the model. The idea is to annihilate the inflow when the queue size overflows the maximum storage capacity. Ideally we would replace $X$ for $\hat{X}$ where:
	\begin{equation*}
		\hat{X}(q(t),t) = H(q(t))\cdot X(t)
	\end{equation*}
	and $H$ is a \textit{Heaviside function}:
	\begin{equation*}
		H(q)=
		\begin{cases}
			1& \text{  for  } q\leq k,\\
			0& \text{  for  } q> k.
		\end{cases}
	\end{equation*} 
	The only problem with this approach is that we would be introducing discontinuities to $q'$ and thus we would not be able to apply existence theorems and standard numerical integration schemes. Hence, we use the \textit{logistic function} as a smooth approximation of $H$:
	\begin{equation*}
		H_{n}(q) = \frac{1}{1 + (\tfrac{1}{H_0}-1)\cdot e^{n(q-k)}}
	\end{equation*}
	where $H_{n}(k) = H_0>0$. Note that:
	\begin{align*}
		\int_{-\infty}^{+\infty}(H(x)-H_n(x))^2dx &= \int_{-\infty}^{k}(1-H_n(x))^2dx + \int_{k}^{+\infty}(H_n(x))^2dx \\
		&= 2\cdot\int_{k}^{+\infty}(H_n(x))^2dx\leq\frac{1}{n}.
	\end{align*}
	Hence $H_n\rightarrow H$ as $n\rightarrow+\infty$ in the $L^2$ norm. Therefore we obtain the following \textit{logistic finite-queue} model:
	\begin{equation}\label{modlogfinite}
		\begin{cases}
			q'(t) &= \hat{X}(q(t),t) - \left[\mu + e^{-\alpha q(t)}\left(\min\{\mu,\hat{X}(q(t),t)\}-\mu\right)\right] \text{  for  } t>t_0,\\
			q(t_0) &= q_0.
		\end{cases}
	\end{equation}
	where 
	\begin{equation*}
		\hat{X}(q(t),t) = H_n(q(t))\cdot X(t).
	\end{equation*}
	\begin{thm}\label{main_thm}
		Let $q(t_0)=q_0\geq 0$ be an initial condition, $X\geq 0$ a continuous and positive inflow function, $\mu>0$ a maximum outflow rate and $k>0$ a maximum queue capacity. Then we have that the system (\ref{modlogfinite}) satisfies the following properties:
		\begin{enumerate}
			\item[1.] There exists a unique and positive solution defined for all times.
			\item[2.] If the inflow is strictly smaller than the maximum outflow rate then the queue gets empty exponentially fast.
			\item[3.] FIFO: the exit time of an entity that has arrived in the instant $t$ is smaller than others that arrive at $s>t$.
			\item[4.] If $q_0<k$ and the inflow satisfies that 
			\begin{equation}
				X(t)\leq M_X \text{  for all  } t\geq t_0 \text{ for some } M_X>0, 
			\end{equation}
			then there exists a $H_0>0$ and a $n>0$ such that $H_n$ approximates $H$ as precisely as desired (in the $L^2$ sense) and moreover 
			\begin{equation*}
				q(t)\leq k \text{  for all  } t\geq t_0.
			\end{equation*}
		\end{enumerate}
	\end{thm}
	\begin{proof}
		Properties 1,2 and 3 follow because the right-hand side of (\ref{modlogfinite}) is \textit{locally Lipschitz continuous} in $q$ so we can apply again the standard existence and uniqueness theorem for ODEs \cite{ODEth}, and afterwards repeat exactly the same arguments given before. Let's prove property number 4. Assume there is a $t'>t_0$ such that $q(t')>k$. Since $q(t_0)<k$, there must exist a $t_*$ such that $q(t_*)=k$. We shall assume that in a sufficiently small interval $I$ around $t_*$ the function $q(t)$ is strictly increasing, so that $q'(t)>0$ for $t\in I$. Analogously as in the proof of Proposition \ref{positividad} we must have that $\hat{X}>\mu$ in $I$. Hence
		\begin{align*}
			q'(t_*) &= \hat{X}(q(t_*),t_*) - \mu = H_0\cdot X(t_*) - \mu \\
			&\leq H_0\cdot M_X - \mu = 0
		\end{align*}
		where we have taken $H_0 = \frac{\mu}{M_X}$. Whence we get a contradiction, so the queue remains bounded by $k$ for all times.
	\end{proof}
	\subsection{{Variable service times}}
	Until this point we have assumed that $\mu$ is constant. Nevertheless it is very easy to see that all results that we have proven are still valid even when $\mu = \mu(t)$ depends continuously on time as long as $\mu(t)\geq0$ for all $t$.
	\subsection{{Multiple servers}}
	The key modification now is to make $\mu = \mu(q(t))$ to depend on the queue size. Assume we have $m$ servers of speed $\mu_0$. We take:
	\begin{equation}
		\mu(q(t))=
		\begin{cases}
			\mu_0m & \text{ if } q(t)\geq m-1 ,\\
			\mu_0\left(1 + q(t)\right) & \text{ if } q(t)\leq m-1.
		\end{cases}
	\end{equation}
	\subsection{{Separation of flows}}\label{sep_flows}
	Assume $X_1$ and $X_2$ are two inflows that join in a system with a server and a queue. They are processed in that system but afterwards they follow different paths. Denote by $Y_1$ and $Y_2$ the corresponding outflows. We can obtain them with a very simple argument. If we denote by $X=X_1+X_2$ and process this inflow, we get an aggregated outflow $Y$. Then for each $t$ we should have that: 
	\begin{equation*}
		\frac{X_1(t)}{X(t)} = \frac{Y_1(t)}{Y(t)},
	\end{equation*}
	hence
	\begin{equation*}
		Y_1(t) = \tfrac{Y(t)}{X(t)}\cdot X_1(t).
	\end{equation*}
	Likewise,
	\begin{equation*}
		Y_2(t) = \tfrac{Y(t)}{X(t)}\cdot Xf_2(t).
	\end{equation*}
	Note that we have that $Y_1(t)+Y_2(t) = Y(t)$ as expected.
	\subsection{{Priority queues}}
	Assume again $X_1$ and $X_2$ are two inflows that join in a system with a server of fixed speed $\mu$. The only difference is that this time we assume that the flow $X_1$ has a priority over $X_2$, i.e. entities coming from $X_1$ will be served first than entities from $X_2$. We will model separately the queues formed by each flow:
	\begin{equation}\label{prioridad}
		\begin{cases}
			q_1'(t) &= X_1(t) - \left[\mu_1 + e^{-\alpha q_1(t)}\left(\min\{\mu_1,X_1(t)\}-\mu_1\right)\right],\\
			q_2'(t) &= X_2(t) - \left[\mu_2 + e^{-\alpha q_2(t)}\left(\min\{\mu_2,X_2(t)\}-\mu_2\right)\right],
		\end{cases}
	\end{equation}
	where we impose that $\mu_1 + \mu_2 = \mu$. The key idea now is to take:
	\begin{equation*}
		\mu_2(q_1(t)) = \frac{X_2(t)}{X(t)}\mu e^{-\alpha q_1(t)}
	\end{equation*}
	where $X=X_1+X_2$. Also we take $\mu_1(q_1(t)) = \mu - \mu_2(q_1(t))$. Observe that:
	\begin{enumerate}
		\item[1.]  If $q_1=0$ then $\mu_1 = \frac{X_1}{X}\mu$ and $\mu_2 = \frac{X_2}{X}\mu$ so each speed is proportional to its flow weight.
		\item[2.]  If $q_1\rightarrow +\infty$ then $\mu_1\rightarrow\mu$ and $\mu_2\rightarrow0$ so just the priority one entities are served and the others are not.
		
	\end{enumerate} 
	
	\section{Evaluation methodology}\label{sec:method}
	
	In this section, we develop the framework we will follow to evaluate the logistic queue model performance. Roughly speaking, we will compare our continuous queue model with a discrete packet simulator. In order to do so, in the following we will explain: how to set up a realistic packet simulator (based on video services), how to aggregate and process the generated video traffic with the logistic queue model and finally what metrics we will use to compare our queue model with the discrete simulator.
	
	\subsection{Video user service characterization}\label{video_user}
	
	In the following we will focus on video service modeling since it is one of the services that occupies more bandwidth in most telecommunication networks and due to its the easiness of characterization (the extension to other kind of services is straightforward). We will model the inflow created by a video user in a discrete way, i.e. packet by packet. For characterizing video streaming, an on-demand video file was served from a set of HTTP servers to a single end-user based on the MPEG-DASH v1.4 standard. On the server-side, two virtual machines each running an Apache HTTP server instance were responsible for serving the audio and the video components, respectively. The video was served at HD 720p and its duration was 10 minutes. 
	
	Now, we will make some statistical assumptions about the packets generated while using a video service:
	\begin{enumerate}
		\item All packets are assumed to be of the same constant size.
		\item Packets are assumed to come in \textit{bursts} of variable size. We will assume that the size of each burst follows a \textit{Normal distribution}.
		\item Bursts are assumed to be separated by variable intervals of time. We will assume that the size of an interval of time separating two bursts --\textit{Interburst time}-- follows an \textit{Exponential distribution}.
		\item Finally, packets are assumed to be separated by variable intervals of time inside a burst. We will assume that the size of an interval of time separating two packets inside a burst --\textit{Interpacket time}-- also follows an \textit{Exponential distribution}.
	\end{enumerate}
	Once the real video flows were generated, the parameters of each distribution were estimated using standard statistical methods. We obtained the results of Table \ref{video_pack_res}.
	\begin{table}[htb!]
		\centering
		\begin{tabularx}{0.66\textwidth}{Xc}
			\toprule
			\textit{Packet size (Bytes)} & 1464 \\ 
			\textit{Burst size (mean)} & 1714 \\ 
			\textit{Burst size (variance)} & 278 \\ 
			\textit{Interburst time (seconds)} & 5.56 \\ 
			\textit{Interpacket time (seconds)} & 0.00345 \\ 
			\bottomrule
		\end{tabularx}
		\caption{Estimation of parameters for a video user.\label{video_pack_res}}
	\end{table}
	
	Next, we want to model the use of video services. Since we are interested in simulating more than a few hours, we cannot assume that each video user is watching videos continuously without stopping. 
	
	\begin{table}[htb!]
		\centering
		\begin{tabularx}{0.66\textwidth}{Xc}		
			\toprule
			\tableheadline{Video use (minutes)} & \tableheadline{Probability} \\ \midrule 
			{5}                                         & 0.4                  \\ 
			{15}                                        & 0.3                  \\ 
			{30}                                        & 0.25                 \\ 
			{120}                                       & 0.05                 \\ 
			\bottomrule
		\end{tabularx}
		\caption{Length of video consumptions and their probabilities.	\label{video_uses} }	
		
	\end{table}
	
	Hence, we assume that \textit{uses of the service} are separated by an interval of time --\textit{Interuse time}-- following an \textit{Exponential distribution} with mean 45 minutes. So on average, a typical user watches some videos every forty-five minutes. Finally, we know from experience that not all videos are of the same length, and even if they were, it is possible and usual to watch more than one. Therefore we also assume a probability distribution on \textit{video consumptions} listed in Table \ref{video_uses}.
	
	\medskip
	
\begin{figure}[h!]
		\begin{center}
			\includegraphics[scale=0.3]{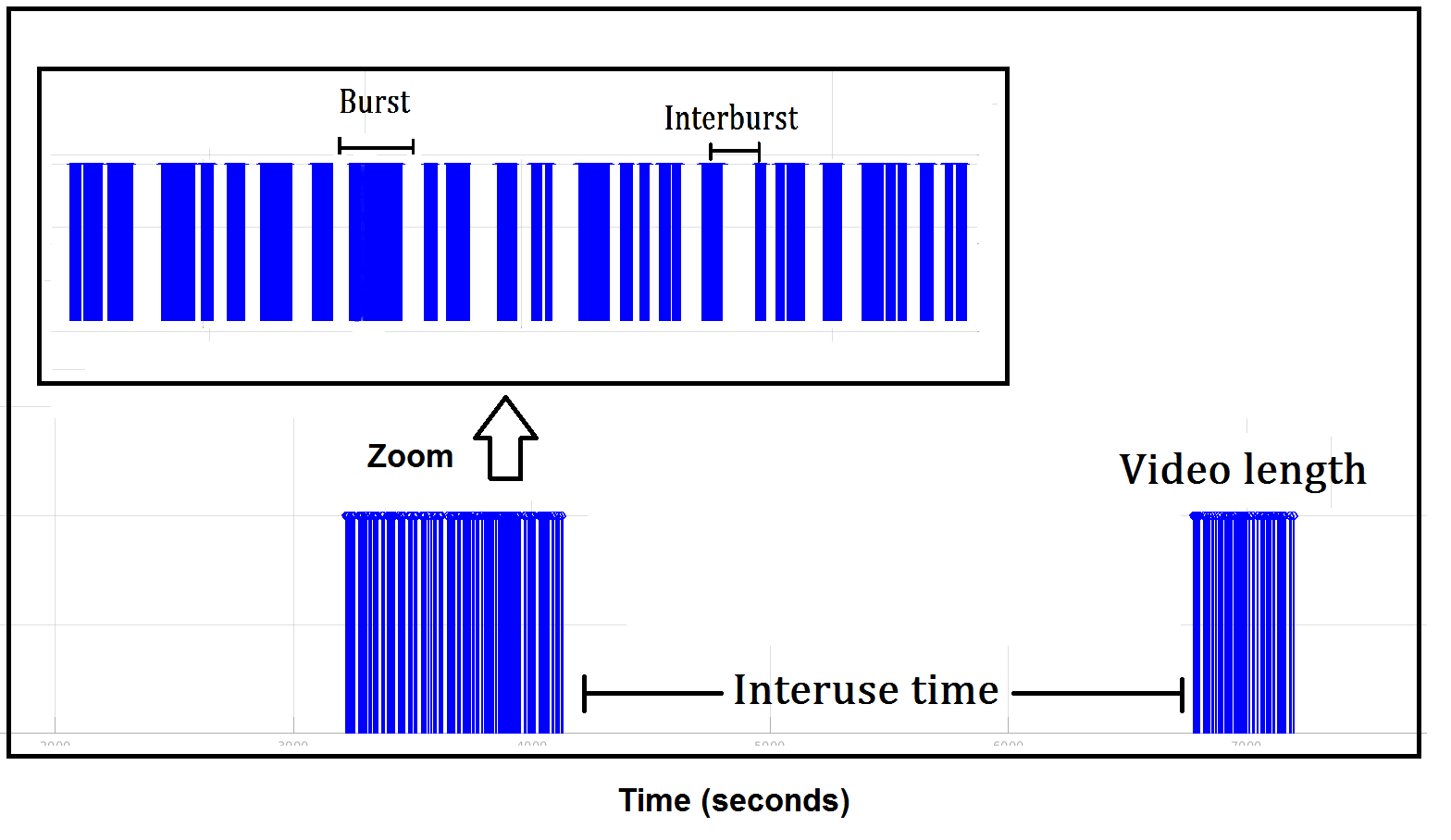}
			\caption{Simulation of packets generated by a video user during two hours. Each blue line represents a packet. \label{paq_disc}}
		\end{center}
	\end{figure}

	Using all previous assumptions we can simulate all packets generated by a video user. This is done by implementing a \textit{discrete packet simulator}  in SIMULINK (a block diagram environment for multidomain simulation and Model-Based Design) using the parameters and statistical distributions just described. In Figure \ref{paq_disc} we show a simulated packet flow using the previously estimated parameters. 
	
	\medskip
	
	Once this input traffic is generated for one user, we can very easily generate and aggregate several video users and serve them through a system with a server of speed $\mu$ and a queue of capacity $k$ (all of this process is performed at the packet level directly in SIMULINK). In such a way we obtain a discrete packet simulator which gives us a very detailed and granular queue size and an outflow which we will use to compare the logistic queue model with. 
	
	\subsection{Numerical integration of the logistic queue model}\label{sec:integration}
	
	At the physical level and in a discrete simulator like the one we just described, data is transmitted discretely in the form of packets using servers from a source node to a destination node in the network. Nevertheless, in contrast to the discrete event simulator, the logistic queue model is a fluid flow model, which means that it does not consider each packet independently but the flow generated by them. Therefore the inflow must be given to the logistic model in a continuous fashion.  In order to do this, we simply count how many entities (packets) have occurred each $dt$ seconds (e.g. $dt = 60$) and then add them up. In more mathematical terms, (e.g. for a video user), in a point of the form $t_{i+1} = (i+1)\cdot dt$ we put
	\begin{equation*}
		X(t_{i+1})\approx\frac{1}{dt}\sum\left(\text{entities occurring in } \left[t_i,t_{i+1}\right]\right).
	\end{equation*}

	Finally, we have to solve the ODE (\ref{modlog}). In order to do that we can use standard mathematical algorithms for the numerical integration of differential equations (i.e. Runge-Kutta methods). The only caution one needs to take care of is the interpolation of the inflow between two successive points $t_i$ and $t_{i+1}$. A linear interpolation suffices. 
	\subsection*{{Queue error due to aggregation time}}
	
	Approximating the inflow as discussed introduces an error in the model that we can not avoid but which we can estimate. Denote by $q_{log}(t)$ the queue size given by the logistic queue model, and by $q_{disc}(t)$ the real queue size (or the one given by the discrete simulator). An entity that enters to the system at $t\in \left[i\cdot dt,(i+1)\cdot dt\right]$ will experience a delay of $q_{disc}(t)/\mu$ seconds, nevertheless aggregating each $dt$ seconds the model can not account the fact that the packet got in at $t$, hence:
	\begin{equation*}
		|\frac{q_{disc}(t)}{\mu}-\frac{q_{log}(t)}{\mu}|\leq dt.
	\end{equation*}
	If the traffic is really intense, i.e. if $\rho = \frac{\lambda}{\mu}$ is near one, the best we can expect is:
	\begin{equation}\label{error_queue}
		|q_{disc}(t)-q_{log}(t)|\sim \mu\cdot (1-\rho)\cdot dt.
	\end{equation}

	\subsection{Error measures}
	In order to compare the queue size given by the logistic queue model, which we denote $q_{log}(t)$, with the queue size given by a discrete packet simulator, which we denote $q_{disc}(t)$ we need some error measures. In simulation those quantities are given at discrete instants of time, so we will only have two vectors $\bf q_{disc},q_{log}$ evaluated at a finite number of times $n$. We will be mostly interested in checking if the model captures big queues since they are the most relevant for telecommunication networks and moreover affect the outflow the most. We define:
	\begin{description}
		\item[1] \textit{Error relative to the maximum}: 
		\begin{equation*}
			\frac{\bf \|q_{disc}-q_{log}\|}{\|\textbf{1}\cdot \max\left(\bf q_{disc}\right)\|}
		\end{equation*}
		where $\textbf{1}$ is vector of ones of length $n$.
		\item[2] \textit{Maximum occupancy error}: 
		\begin{equation*}
			\frac{|\max\left(\bf q_{disc}\right)-\max\left(\bf q_{log}\right)|}{|\max\left(\bf q_{disc}\right)|}.
		\end{equation*}
	\end{description}
	Likewise, we will want to compare the outflow given by the logistic queue model, which we denote $Y_{log}(t)$, with the outflow given by the discrete packet simulator, which we denote $Y_{disc}(t)$. We obtain two vectors $\bf Y_{disc} \it = (Y_{disc}^1, \ldots,Y_{disc}^n)$ and $\bf Y_{log} \it = (Y_{log}^1, \ldots,Y_{log}^n)$ evaluated at a finite number of times $n$. We define:
	\begin{description}
		\item[3] \textit{Mean relative outflow error}: 
		\begin{equation*}
			\frac{1}{n}\sum_{i=1}^{n}\frac{|Y_{disc}^i-Y_{log}^i|}{| Y_{disc}^i|}
		\end{equation*}
		
		\item[4] \textit{Global relative error}: 
		\begin{equation*}
			\frac{\bf \|Y_{disc}-Y_{log}\|}{\|\bf Y_{disc}\|}
		\end{equation*}
		
	\end{description}

	\section{Validations and results}\label{sec:validation}
	
	In this section we validate the logistic queue model comparing its performance with the discrete event simulator described in the previous section. As already explained, our goal is to compare the queue size given by the logistic queue model with the queue size given by a discrete packet simulator. We implemented both models using standard software: \textit{the logistic queue model} was implemented in MATLAB. As already explained, the only thing that had to be done was to solve the ordinary differential equation (\ref{modlog}). In Matlab, there are several libraries for solving differential equations numerically, being the most usual ones \textit{ode45} and \textit{ode113}. Both functions give indistinguishable results. The \textit{discrete packet simulator} on the other hand was implemented in SIMULINK using the parameters described in the previous section for the traffic generation of video users. All comparisons are performed using an Intel Core i5-12400F with 6 cores of 2.5 GHz.

	\subsection{Illustrative example}
	In order to have an intuitive understanding of the model and all the things involved with it, we will first give an illustrative example of its performance. Later we will make a more exhaustive performance analysis. We assume the following particular scenario:
	\begin{enumerate}
		\item We aggregate 10 video users in a server of maximum velocity $\mu = 11.33$ Mb/s during 2 days.
		\item The queue is assumed to be large enough so that we do not lose any packets.
	\end{enumerate}
	For generating the video flows we use the method described in Section \ref{video_user}. The goal is to predict the queue formed under this scenario and the outflow we get. 
	\subsubsection{Inflow}
\begin{figure}[h!]
		\begin{center}
			\includegraphics[scale=0.8]{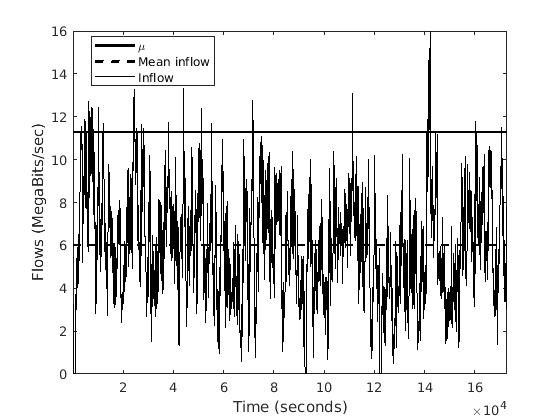}
			\caption{Inflow to the system. Its standard deviation is 2.53 Mb/s.\label{inflow_ex}}
		\end{center}
	\end{figure}

	Adding 10 video flows $X_v$ we get the inflow of Figure \ref{inflow_ex}. In other words we put
	\begin{equation*}
		X(t) = \sum_{v=1}^{10}X_v(t).
	\end{equation*}
	
	\noindent The mean of the inflow is $\lambda = 6$ Mb/s, hence the average intensity is $\rho = 53\%$. Therefore we take $\alpha = \tfrac{\rho}{\mu}=0.05$. As we see in the picture, during some periods of time, the inflow is bigger than the maximum outflow $\mu$, therefore in those moments, we expect the queue size to increase. 
	
	\subsubsection{Comparison of queues}
\begin{figure}[h!]
		\begin{center}
			\includegraphics[scale=0.85]{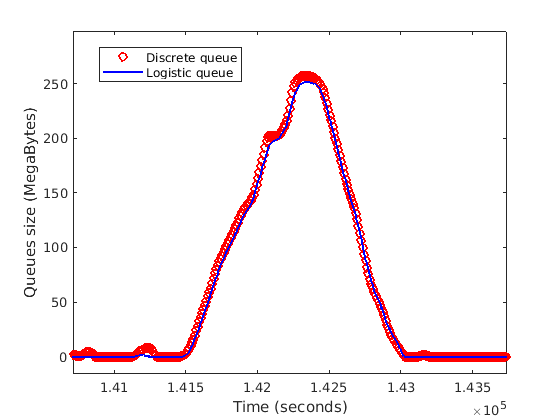}
			\caption{Superposition of logistic and discrete queues in a neighborhood of the maximum queue occupancy.\label{colas_superpuestas}}
		\end{center}
	\end{figure}

	We feed the inflow $X$ to both the discrete event simulator and to the logistic queue model and compare the queue results we get in both cases. Note that $X$ is given to the discrete simulator in discrete form, i.e. packet by packet as it was generated, whereas to the logistic model, the inflow is given in a continuous fashion, i.e. aggregated every 60 seconds and then linearly interpolated when needed.
	\begin{figure}[h!]
		\centering
		\begin{minipage}{.5\textwidth}
			\centering
			\includegraphics[scale=.5]{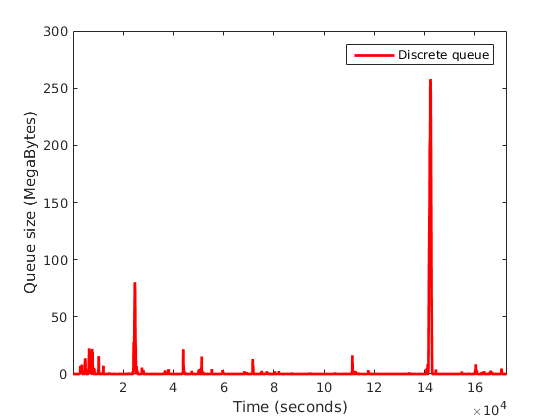}
			\caption{Queue of the discrete simulator.\label{fig:test1}}
			
		\end{minipage}%
		\begin{minipage}{.5\textwidth}
			\centering
			\includegraphics[scale=.5]{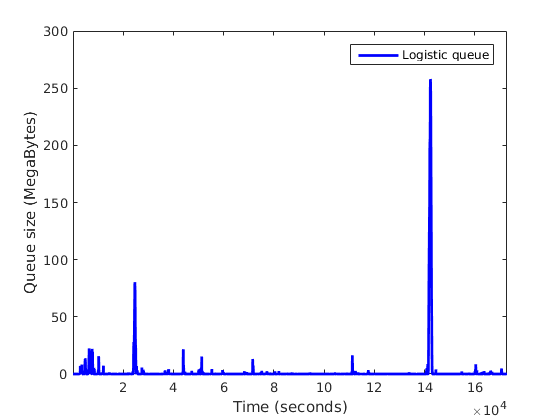}
			\caption{Queue of the logistic model.\label{fig:test2}}
			
		\end{minipage}
	\end{figure}
	Just looking at Figures \ref{fig:test1} and \ref{fig:test2} of the queues formed is difficult to tell the difference, hence we use the two error measures introduced before:
	\begin{enumerate}
		\item[-] \textit{Error relative to the maximum}: 
		\begin{equation*}
			\frac{\bf \|q_{disc}-q_{log}\|}{\|\textbf{1}\cdot \max\left(\bf  q_{disc}\right)\|} = 0.63\%
		\end{equation*}
		\item[-] \textit{Maximum occupancy error}: 
		\begin{equation*}
			\frac{|\max\left(\bf q_{disc}\right)-\max\left(\bf q_{log}\right)|}{|\max\left(\bf q_{disc}\right)|} = 2.08\%
		\end{equation*}
	\end{enumerate}
	In Figure \ref{colas_superpuestas} we examine both queues superposed in a neighborhood of the maximum occupancy.

	\subsubsection{Queue error due to aggregation time}
	Recall that if the traffic is intense, i.e. if $\rho = \frac{\lambda}{\mu}$ is near one, the best we can expect is:
	\begin{equation*}
		\left|\frac{q_{disc}(t)}{\mu}-\frac{q_{log}(t)}{\mu}\right|\sim (1-\rho)\cdot dt.
	\end{equation*}
	In our example, we actually have that
	\begin{equation*}
		\max_{t\in[t_0,t_1]} \left|\frac{q_{disc}(t)}{\mu}-\frac{q_{log}(t)}{\mu}\right| = 15.71 \text{ seconds},
	\end{equation*}
	so even the estimate with $(1-\rho)\cdot dt = 28.18 $ seconds holds.
	
	\subsubsection{Comparison of outflows}
	We can also compare both outflows, the one given by the discrete simulator and the other computed from the logistic queue. 
	\begin{figure}[h!]
		\centering
		\begin{minipage}{.5\textwidth}
			\centering
			\includegraphics[scale=.5]{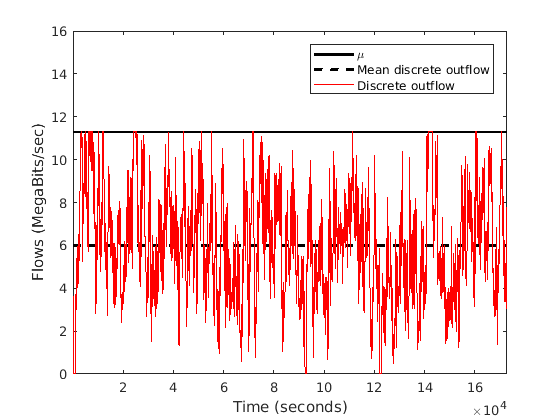}
			\caption{Outflow of discrete simulator.\label{out_disc}}
		\end{minipage}%
		\begin{minipage}{.5\textwidth}
			\centering
			\includegraphics[scale=.5]{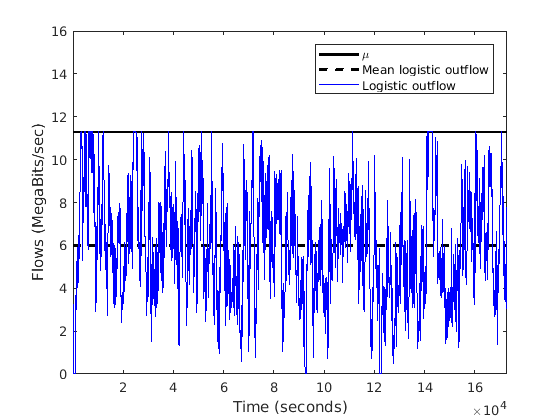}
			\caption{Outflow of logistic model.\label{out_log}}
		\end{minipage}
	\end{figure}
	As expected both flows do not exceed the maximum capacity $\mu$. Looking at Figures \ref{out_disc} and \ref{out_log} is again very difficult to tell the difference. We again use the two error metrics already introduced:
	\begin{enumerate}
		\item[-] \textit{Mean relative outflow error}: 
		\begin{equation*}
			\frac{1}{n}\sum_{i=1}^{n}\frac{|Y_{disc}^i-Y_{log}^i|}{| Y_{disc}^i|} = 0.67\%
		\end{equation*}
		
		\item[-] \textit{Global relative error}: 
		\begin{equation*}
			\frac{\bf \|Y_{disc}-Y_{log}\|}{\|\bf Y_{disc}\|} = 1.58\%
		\end{equation*}
	\end{enumerate}
	
	\subsubsection{Comparison with no queue model}
	In order to have a reference to compare with, we can compute the previous error measures with the inflow to the system. This would be the simplest model in which we assume that the outflow equals the inflow and the queue has no effect whatsoever. In this example, the mean relative error is: 
	\begin{equation*}
		\frac{1}{n}\sum_{i=1}^{n}\frac{|Y_{disc}^i-X^i|}{| Y_{disc}^i|} = 0.75\%, 
	\end{equation*}
	while the global relative error is:
	\begin{equation*}
		\frac{\bf \|Y_{disc}-X\|}{\|\bf Y_{disc}\|} = 5.27\%.
	\end{equation*}
	As expected mean-wise the relative errors are very similar. This is due to the fact that most of the time the inflow is smaller than $\mu$ and hence the outflow $g$ is equal to $f$. On the other hand, the global relative error is sensibly bigger for the no queue model since this measure penalizes more point-wise errors. 
	
	\subsubsection{Discussion of the example}
	As already stated the goal of this example was to introduce in an intuitive fashion all concepts involved with the model in a practical setting. The results show that our model captures very well the queue generated and hence also the outflow. At the end of the day, one may wonder why it is worth using a continuous model (the logistic queue model) if we already can make discrete simulations and get quite accurate results. The key is of course \textit{scalability}. In this illustrative example, the simulation time of the discrete simulator was around 60021 seconds, which is about 16 hours. On the other hand, using the logistic queue model, the whole process only took about 29 seconds. Hence the continuous model is around 2000 times faster. This is expectable since the discrete simulator had to process around $2.45\cdot10^7$ entities. In the next section, we will make a more comprehensive analysis of the performance of the logistic queue model.    
	
	\subsection{Permormance Analysis}
	In this subsection, we analyze how the error measures introduced in the previous examples evolve as the intensity of the inflow is varied. Again we assume the same scenario as in the previous section: 
	
	\begin{enumerate}
		\item We aggregate 10 video users in a server of maximum velocity $\mu = 11.33$ Mb/s during 2 days each 60 seconds.
		\item The queue is assumed to be large enough so that we do not lose any packets.
	\end{enumerate}
	We vary inflow intensities $\rho$ from approximately $45\%$ to $85\%$. In order to do so we simply reduce the \textit{Interuse time} introduced before when modeling a video user in Section \ref{video_user}. Recall that this parameter represented the amount of time between two uses of video service. Hence reducing it, we only make our 10 users watch videos more often. In total 21 simulations were launched both in the discrete event simulator implemented in SIMULINK and using the logistic queue model implemented in MATLAB. Of the 21 simulations 3 of them had not finished after more than a month in the discrete simulator, and therefore they were discarded. Note that in each simulation we are generating around 25 million packets, so in total in 21 simulations we generated about 500 million packets.
	
	\subsubsection{{Scalability: simulation times}}
	We first compare the simulation time of both models. Of the 18 simulations analyzed, the mean simulation time of the logistic model is 26.3 seconds whereas the mean of the discrete simulator is 183.01 hours (almost 8 days). As seen in Figure \ref{simulation_times}, the logistic queue model is about 8 orders of magnitude faster than the discrete simulator.
	
	\subsubsection{{Comparison of queues}}
\begin{figure}
		\begin{minipage}[c]{0.475\linewidth}
			\includegraphics[width=\linewidth]{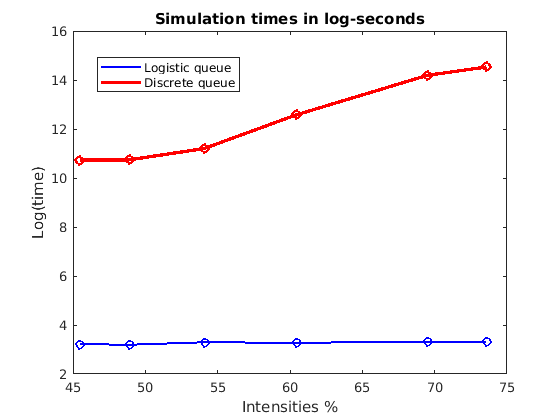}
			\caption{Simulation times of logistic and discrete queue model varying the inflow intensity.\label{simulation_times}}
		\end{minipage}
		\hfill
		\begin{minipage}[c]{0.475\linewidth}
			\includegraphics[width=\linewidth]{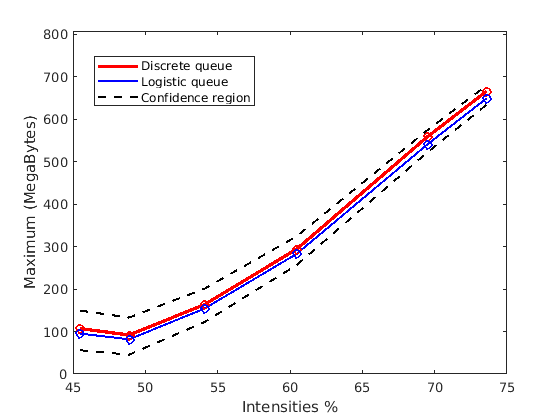}
			\caption{Comparison of maximum queue size for both models varying the inflow intensity. \label{max_queues}}
			
		\end{minipage}%
	\end{figure}

	We now compare the maximum queue size given by both models. In Figure \ref{max_queues} bellow we can see a plot of both curves for different intensities. As we see both curves are quite close. In dashed lines, we have plotted the error bound (\ref{error_queue}). 
	
	\begin{figure}[hbt!]
		
		\begin{subfigure}{.475\linewidth}
			\includegraphics[width=\linewidth]{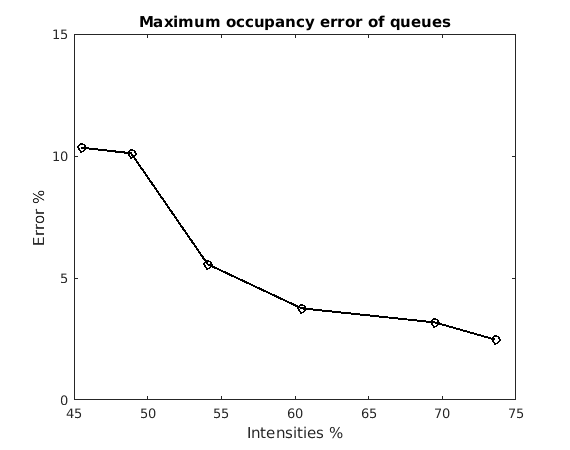}
			\caption{Maximum occupancy error of the logistic model with respect to the discrete simulator varying the inflow intensity.\label{max_ocup_error}}
		\end{subfigure}\hfill 
		\begin{subfigure}{.475\linewidth}
			\includegraphics[width=\linewidth]{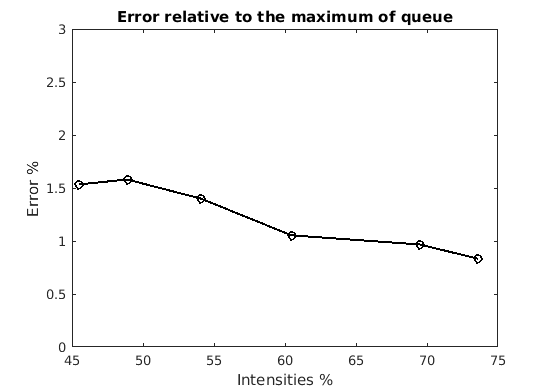}
			\caption{Error relative to the maximum of the logistic model with respect to the discrete simulator varying the inflow intensity.\label{err_rel_mx}}
		\end{subfigure}
		
		\medskip 
		\begin{subfigure}{.475\linewidth}
			\includegraphics[width=\linewidth]{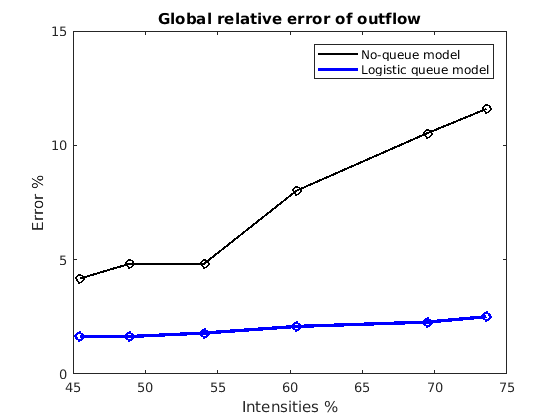}
			\caption{Global relative error of the outflows given by logistic model and the no-queue model with respect to the discrete simulator varying the inflow intensity.\label{glob_rel_error}}
		\end{subfigure}\hfill 
		\begin{subfigure}{.475\linewidth}
			\includegraphics[width=\linewidth]{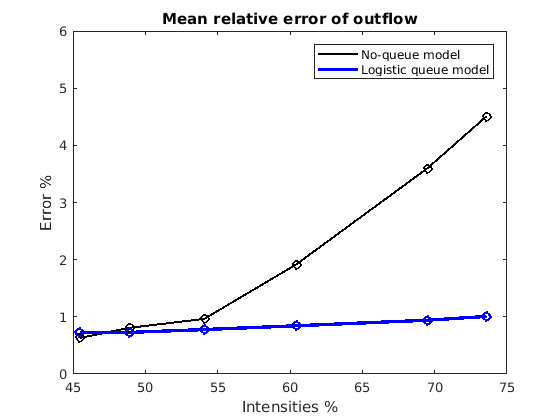}
			\caption{Mean relative error of the outflows given by logistic model and the no-queue model with respect to the discrete simulator varying the inflow intensity.\label{mean_rel_error}}
		\end{subfigure}
		
		\caption{Error measures comparing the logistic queue model with the discrete packet simulator. }
		\label{fig:roc}
	\end{figure}
	
	It is also interesting to see how the maximum occupancy error evolves as the intensity increases. We see in Figure \ref{max_ocup_error} how the error decreases as the intensity increases.
	In Figure \ref{err_rel_mx} we plot the error relative to the maximum introduced earlier. Again the error reduces as the intensity increases. This is again in accordance with (\ref{error_queue}).
	
	\subsubsection{{Comparison of outflows}}
	Finally, we compare the outflow given by the logistic queue model with the one given by the discrete simulator. To have a reference to compare with we also plot the error of the no-queue model as discussed before. Recall that this just assumes that the queue has no effect whatsoever so the outflows just equals the inflow. We begin by plotting the global relative error in Figure \ref{glob_rel_error}.
	
	Note how the error of the no-queue model increases as the intensity increases. This is to be expected because increasing the intensity increases the queue size and therefore makes the outflow differ more from the inflow. On the other hand, the error of the logistic queue model seems quite stable, its mean is around $2\%$. 
	
	Finally, we compute the mean relative error. In Figure \ref{mean_rel_error} we observe that until $\rho = 55\%$ there is no big difference between both models, both errors are quite low. This is again logical since we know that the outflow of the system is almost identical to the inflow when there are not a big queues.

	\section{Digital Twin (DT) use case}\label{sec:scenario}
	
	In this section, we present a DT use case using the logistic queue model in a more complex communication network. Basically, we study how the latency of the network is affected when we introduce an external priority flow into play. We will be dealing with flows of the order of Gb/s, this is where having an efficient model becomes of critical importance since running a discrete simulation for these orders of magnitude would be infeasible. 
	
	\subsection{Scenario}
	
\begin{figure}[H]
		\begin{center}
			\includegraphics[width=\linewidth]{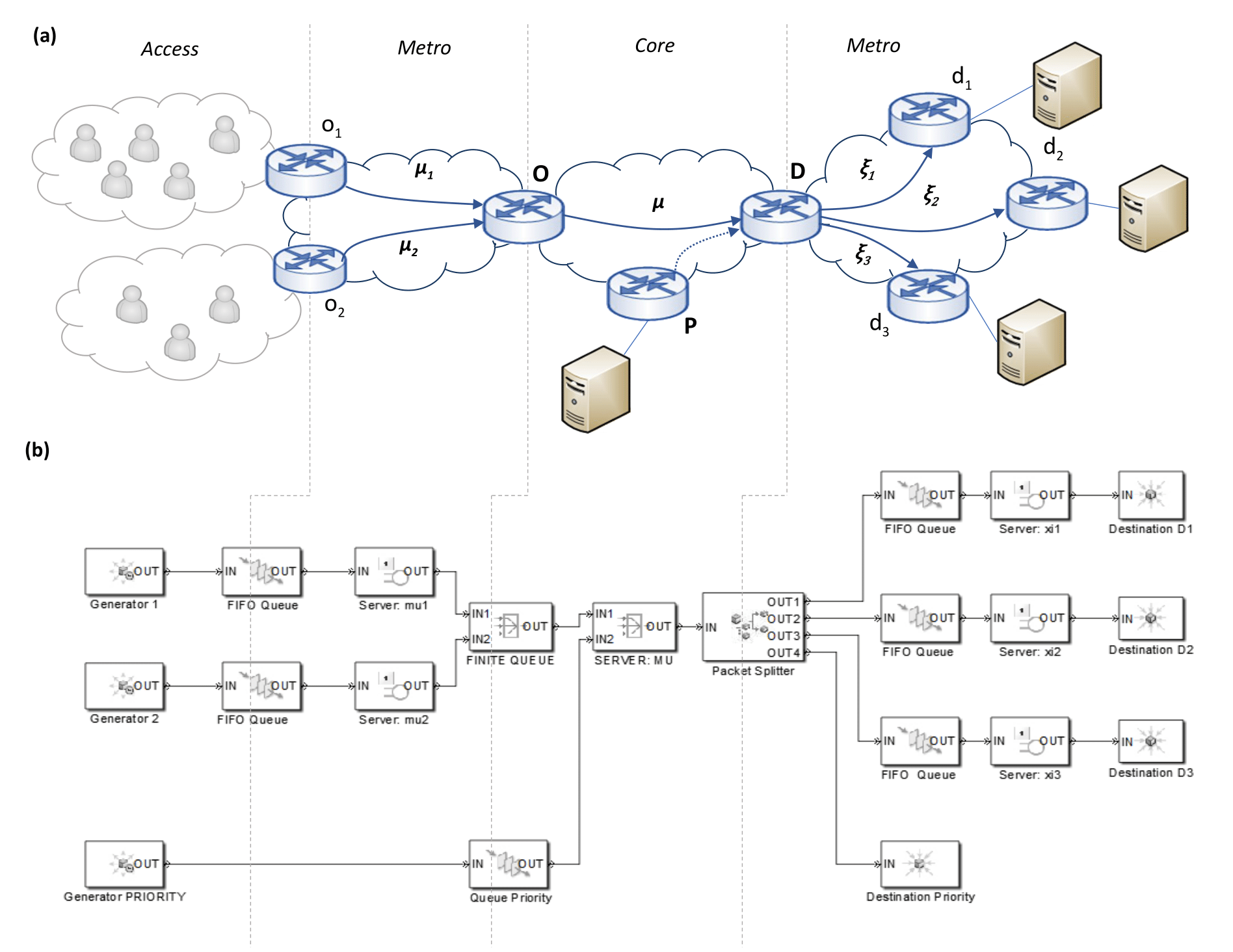}
			\caption{Network example considered as real network (a) and the DT built using the logistic queue model as the main building block (b). The goal is to send packets from region $o$ to region $d$ of the network. \label{dibujito}}
		\end{center}
	\end{figure}

	Figure \ref{dibujito} (a) presents the network example considered as the real network, whereas Figure \ref{dibujito} (b) shows the DT built using the logistic queue model as the main building block. The example comprises a classical network operator scenario where end users in access networks want to access some services running in remote data centers, which requires connectivity across intermediate metro and core networks \cite{beyond5G}.
	
	More specifically, the network scenario considers the following assumptions:

	\begin{enumerate}
		\item  A set of flows $X_1,\ldots,X_n$ are generated from secondary packet nodes $o_i$ of an optical network. We assume we have $n$ nodes $o_i$.
		
		\item A proportion $p_{ij}$ of each flow $X_i$ has to be sent from $o_i$ to a specific packet node $d_j$ at the other side of the network. We assume we have $m$ nodes $d_j$.
		
		\item This is done through a high bandwidth connection from an origin node $O$ to a destination node $D$. This link has a server of speed $\mu$ with a queue attached to it of maximum capacity $k$.

		\item Each flow $X_i$ travels through a link between the node $o_i$ and the node $O$. That given connection has a server of velocity $\mu_i$ and a queue. These queues are assumed to be large enough so that we do not lose any packets.

		\item Finally we assume that links between the node $D$ and the destination nodes $d_j$ have a server of velocity $\xi_j$ and a queue. Likewise, these queues are assumed to be large enough so that we do not lose any packets.

	\end{enumerate}

	We will be mainly interested in computing the \textbf{latency} $L_{ij}(t)$ between node $o_i$ and node $d_j$ at instant $t$. Physically this represents the amount of time it takes for a packet to travel from $o_i$ to $d_j$ if it departures at $t$. In our system, we have that:
	\begin{equation*}
		L_{ij}(t) = \frac{q_i^o(t)+s }{\mu_i} + \frac{ q(t_o) + s }{\mu} + \frac{ q_j^d(t_d) + s }{\xi_j},
	\end{equation*}
	
	where
	\begin{description}
		\item[-] $s$ is the size of each packet.
		\item[-] $ q_i^o(t)$ is the queue formed between node $o_i$ and $O$ at $t$.
		\item[-] $ q(t_o)$ is the queue formed between node $O$ and $D$ at $t_o$.
		\item[-] $t_o$ is the instant at which the packet arrives to $O$: $t_o = t + \frac{q_i^o(t)+s }{\mu_i}$.
		\item[-] $ q_j^d(t_d)$ is the queue formed between node $D$ and $d_j$ at $t_d$.
		
		\item[-] $t_d$ is the instant at which the packet arrives to $D$: $t_d = t_o + \frac{q(t_o)+s }{\mu_i}$.
	\end{description}
	
	It is easy to see that the previous formula can be generalized for more general networks. Moreover we define the \textbf{expected latency} of going from region $o$ of the network to region $d$ as:
	
	\begin{equation*}
		L_{od}(t) = \frac{1}{n\cdot m}\sum_{i,j}L_{ij}(t).
	\end{equation*}
	
	Finally, we define the \textbf{ maximum expected latency} as 
	
	\begin{equation*}
		L_{max} = \max_{t} L_{od}(t).
	\end{equation*}
	
	We will be interested in studying how this quantity changes when we modify the scenario just described above.
	
	\subsection{Additional considerations}
	
	Applying our queue model in the scenario just described is quite straightforward. First, we process each flow $X_i$ using the logistic queue model (\ref{modlog}) to obtain flows $Y_i$. These are the outflows of each $o_i$ and the inflows to $O$. Doing this we obtain the functions $q_i^o(t)$. Then we put $Y:=\sum_{i=1}^{n}Y_i$ and process this flow using the logistic finite queue model (\ref{modlogfinite}) to obtain a new flow $Z$. This is the outflow of $O$ and the inflow to $D$. Doing this we obtain the function $q(t)$. Now, we must take into account the origin $o_i$ of each flow and its destination $d_j$, hence as explained in Section \ref{sep_flows}:
	\begin{equation*}
		Z_j(t) = \sum_{i=1}^{n}p_{ij}\cdot\left(\frac{Y_i(t)}{Y(t)}\cdot Z(t)\right).
	\end{equation*}
	
	This is the outflow of $D$ and the inflow to $d_j$. Finally, we process each $Z_j$ using again the the logistic queue model (\ref{modlog}). Doing this we obtain the functions $q_j^d(t)$.
	\subsection{Results}
	\begin{description}
		\item[-] For generating the flows $X_i$ we use the method described in Section \ref{video_user} when we modeled a video user. In total $n=4$ flows were generated during 1 day. 
		
		\item[-] Each flow $X_i$ consists of 10,000 video users, the mean inflow rate of each flow is about 12.5 Gb/s. We take all $\mu_i = 25 $ Gb/s.
		
		\item[-] For the big link OD we take $\mu = 100$ Gb/s and the maximum capacity $k = 25$ GigaBytes. 
		
		\item[-] Since the mean of each flow $X_i$ is about 12.5 Gb/s and $n=4$, we expect the use of the OD link to be around $50\%$.
		
		\item[-] We take $m=5$ nodes $d_j$. Each link has a speed of $\xi_j = 20 $ Gb/s.
		
		\item[-]  Finally, the matrix $p_{ij}$ is chosen to be:
		\[
		p=
		\left[ {\begin{array}{ccccc}
				0.1293 &   0.3124 &   0.0548 &   0.2534 &   0.2501\\
				0.1600 &   0.1681 &   0.0497 &   0.2203 &   0.4019\\
				0.3029 &   0.0009 &   0.1687 &   0.2224 &   0.3051\\
				0.0042 &   0.3710 &   0.2344 &   0.0250 &   0.3655\\
		\end{array} } \right],
		\]
		\noindent so, for instance, $12.93 \%$ of the packets generated from node $o_1$ go to node $d_1$.
	\end{description}
	
\begin{figure}
		\begin{minipage}[c]{0.45\linewidth}
			\includegraphics[width=\linewidth]{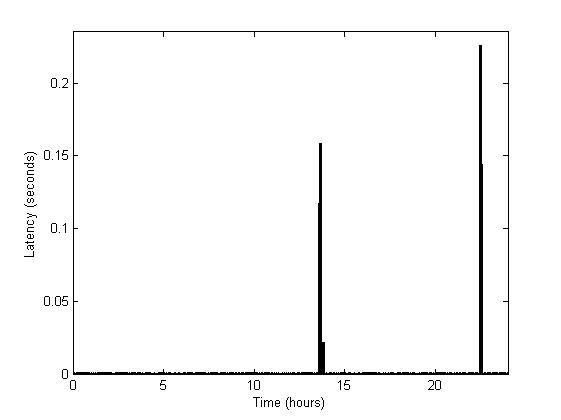}
			\caption{Expected latency in seconds of going from region $o$ to region $d$. \label{latencia0}}
		\end{minipage}
		\hfill
		\begin{minipage}[c]{0.45\linewidth}
			\includegraphics[width=\linewidth]{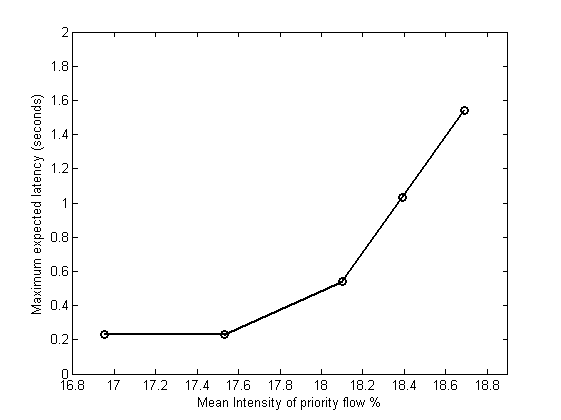}
			\caption{Maximum latency going from region $o$ to region $d$ varying the intensity of an extra priority flow. \label{latencia2}}
		\end{minipage}%
	\end{figure}

	Then, in this setting, we get the expected latency of Figure \ref{latencia0}. We see that for a packet that departs from region $o$ of the network, we expect that it takes at most $0.2$ seconds to get to its destination in region $d$. The whole simulation (1 day of simulated time) took only about 2 minutes to be completed.
	
	\bigskip
	
	Now, assume that another flow of packets needs to be sent to $D$ using the same link $OD$ that the node $O$ was using (represented in Figure \ref{dibujito} by the node $P$). We will assume that the flow going from $P$ to $D$ has priority over the flow going from $O$ to $D$. We implement this using model (\ref{prioridad}): we inject an additional priority flow into the system. We are interested in seeing how this will affect the maximum expected latency of going from region $o$ to region $d$. The results, varying the intensity of the priority flow are shown in Figure \ref{latencia2}.

	As expected the latency gets bigger as the priority flow gets bigger. Nevertheless, it is interesting to observe that for priority flows smaller than $18$ Gb/s we get a tolerable maximum expected latency of about 0.5 seconds. Afterwards, the increase is exponential and we get unacceptable latencies.

	\section{Conclusions}
	In this work, we have presented for the first time the formal derivation and mathematical properties of the \textit{logistic queue model} that first appeared in \cite{Ru18}. We have proven mathematically that the model has all the desirable theoretical properties one should expect: (i) positivity of the queue size, (ii) well-behaved asymptotic behavior: the queue gets empty if the inflow does not overflow, and (iii) FIFO property: the system satisfies a first-in first-out discipline. Moreover, in contrast with the famous Vickrey's point-queue model, the logistic queue model allows us to easily explore multiple extensions to more general scenarios such as finite queues (see Theorem \ref{main_thm}), multiple servers, priority queues, etc.
	
	\medskip
	
	We validated our queue model comparing it with a discrete event simulator. We showed that in terms of queue size and outflow prediction, our model is as precise as a discrete one, with relative errors in maximum queue sizes of the order of $1$-$2\%$ with the advantage of speed in simulations. We compared simulation times and concluded that the logistic queue model is around 7 or 8 orders of magnitude faster than a discrete one for the scenarios analyzed. 
	
	\medskip
	
	Finally, we applied the proposed logistic queue model to build the DT of a communication network and computed the expected latency in the event of different traffic configurations. The results show the applicability of our model to compute KPIs in DT environments, which is a key feature that autonomous network control systems need to provide in order to efficiently manage next-generation communication networks.
	
	
	
	
	
	\section{Acknowledgments}
	This research has received funding from the NextGeneration UNICO5G TIMING (TSI-063000-2021-145), the SNS JU though the European Union’s Horizon Europe under G.A. No. 101095890 (PREDICT-6G), the AEI IBON (PID2020-114135RB-I00) projects, and the ICREA institution. Moreover F. Coltraro is supported by CSIC project 202350E080 and SGR RobIRI 2021 SGR 00514.

\end{document}